\documentclass[preprint,12pt]{elsarticle}

\usepackage{siunitx}  
\usepackage{graphicx}
\usepackage{subfigure}
\usepackage{amsmath}
\usepackage{xcolor}

\usepackage{geometry}
\usepackage[hidelinks]{hyperref}
\usepackage[all]{hypcap}
\usepackage[noabbrev, capitalise]{cleveref}
\usepackage{amssymb}

\makeatletter
\def\ps@pprintTitle{%
 \let\@oddhead\@empty
 \let\@evenhead\@empty
 \def\@oddfoot{\centerline{\thepage}}%
 \let\@evenfoot\@oddfoot}
\makeatother

\begin{document}

\begin{frontmatter}



\title{Diamagnetically levitating resonant weighing scale}


\author[1]{Xianfeng Chen\corref{cor1}}
\author[1]{Nimit Kothari}
\author[1]{Ata Keşkekler}
\author[1,2]{Peter G. Steeneken}
\author[1]{Farbod Alijani\corref{cor2}}

\address[1]{Department of Precision and Microsystems Engineering, Delft University of Technology, Delft, The Netherlands.}
\address[2]{Kavli Institute of Nanoscience, Delft University of Technology, Lorentzweg 1, 2628 CJ, Delft, The Netherlands.}
\cortext[cor1]{x.chen-7@tudelft.nl}
\cortext[cor2]{f.alijani@tudelft.nl}

\begin{abstract}
Diamagnetic levitation offers stable confinement of an object from its environment at zero power, and thus is a promising technique for developing next generation unclamped resonant sensors. In this work, we realize a resonant weighing scale using a graphite plate that is diamagnetically levitating over a checkerboard arrangement of permanent magnets. We characterize the bending vibrations of the levitating object using laser Doppler vibrometry and use microgram glass beads to calibrate the responsivity of the sensor's resonance frequency to mass changes. The sensor is used for real-time measurement of the evaporation rate of nano-litre droplets with high-accuracy.  By analyzing the resonator's frequency stability, we show that the millimeter graphite sensor can reach mass resolutions down to $\SI{4.0}{ng}$, relevant to biological and chemical sensing concepts. 
\end{abstract}




\end{frontmatter}


\section{Introduction}
\label{sec: introduction}
Mechanical resonators are nowadays being adopted in billions of products, including quartz crystals, Micro-Electro-Mechanical Systems (MEMS), and acoustic wave resonators for time-keeping, frequency referencing and electronic filtering, but also for addressing a wide range of sensor applications in modern technology. These resonant sensors can be used to measure parameters like mass \cite{jensen2008atomic,chaste2012nanomechanical,yang2006zeptogram}, stiffness \cite{zhang2014detecting}, density \cite{belardinelli2020second}, viscosity \cite{wingqvist2007shear}, and pressure \cite{morten1992resonant}, for a diverse range of applications, ranging from environmental monitoring to life sciences \cite{waggoner2007micro,johnson2012biosensing,cooper2009label}.  The working principle of these devices is based on tracking a resonance frequency $f_\mathrm{r}$ of the sensor that depends on the sensing parameter. In this framework, the minimum resonance frequency change that can be detected by the sensor, its limit of detection, depends both on its responsivity and on the uncertainty $\sigma_a$ in the resonance frequency measurement. For low uncertainty or low operation power, a high mechanical quality factor $Q$ of the resonance is beneficial. However, $Q$ can be substantially degraded through clamping, friction, adhesion and aerodynamic losses \cite{schmid2016fundamentals}.

In literature great efforts have been undertaken to reduce these dissipation mechanisms in resonant sensors. For instance, optimizing clamping points by high tension tethers \cite{norte2016mechanical}, or using bulk acoustic modes \cite{wingqvist2007shear}, acoustic Bragg mirrors \cite{wang2011acoustic}, and phononic crystals \cite{ghadimi2018elastic}, have been used for boosting $Q$ of resonant sensors and isolating them from their environment. Ultimate confinement of a sensor though can be obtained by levitating it, which therefore has the potential to significantly improve $Q$ of resonant sensors.

Levitation of objects can be realized in different ways, using optics, acoustics, or magnetic forces \cite{brandt1989levitation}. Among them, diamagnetic levitation stands out as the only method that obtains stable levitation at room temperature without power consumption \cite{d92503966a0041f9a349abdf1003f35a}.  Such passive levitation is important in many practical, high-volume microscale applications, not only because it is difficult to guarantee a continuous power supply for levitation, but also because collapse of a levitating microstructure in the absence of power leads to failure by adhesion. Additionally, unlike optical levitation \cite{gieseler2012subkelvin}, passive levitation does not dissipate power that can heat the levitating object. Interestingly, many materials are diamagnetic, exhibiting negative magnetic susceptibility  \cite{simon2000diamagnetic}, thus making diamagnetic levitation potentially a widely applicable method, especially at the microscale where larger magnetic field gradients can be achieved such that also weakly diamagnetic materials can be levitated. During the last decade, liquid droplets \cite{chetouani2006diamagnetic} and small solid particles \cite{pigot2008diamagnetic} have been levitated stably by micro magnets. In addition, diamagnetic levitation has shown the potential for realizing accelerometers \cite{garmire2007diamagnetically}, energy harvesters \cite{palagummi2018bi}, density \cite{gao2019centrifugal}, and force sensors  \cite{li2006lateral}.

In this paper, we propose a resonant mass sensor based on diamagnetic levitation. The sensor comprises a pyrolytic graphite plate that is passively levitating over a checkerboard arrangement of permanent magnets. Compared to conventional clamped resonant mass sensors, the levitating sensor offers contactless and free motion of the resonator in the absence of clamping and friction forces, potentially leading to higher $Q$ and thus more sensitive sensor design. We characterize the first 10 bending modes of the levitating plate in the kHz regime using a Polytec laser Doppler vibrometer (LDV), and calibrate its mass responsivity using glass microbeads. With the calibrated plate we measure the mass evaporation rate of small liquid droplets on the plate, which is found to be in good agreement with estimates made using the droplet volume. Finally, we characterize the Allan deviation of the resonance frequency $\sigma_a$ and use it to show that the mass resolution of the weighing scale is as low as a few nano-grams (ng). 





\section{Mass sensing working principle}
The working principle of a resonant mass sensor is based on the frequency shift $\delta f$ of the resonant mode when a mass $\delta m$ is attached to the resonator. The dynamics of the bare plate can be described by an effective stiffness $K_\mathrm{eff}$ and an effective mass $M_\mathrm{eff,plate}$. When a small mass $\delta m$ is placed at the anti-node, the effective mass becomes $M_\mathrm{eff}=M_\mathrm{eff,plate}+\delta m$, and the resonance frequency can be written as  $f_{\mathrm{r}}= \frac{1}{2\pi}\sqrt{\frac{K_\mathrm{eff}}{M_\mathrm{eff}}}$. Assuming that the added mass does not influence $K_\mathrm{eff}$ and the quality factor $Q$ of the resonator, the following relation between $\delta f$ and $\delta m$ can be obtained by differentiating the resonance frequency $f_{\mathrm{r}}$ with respect to $\delta m$:
\begin{equation}
    \centering
    \label{eq:mass sensing}
    \delta m=\frac{1}{\mathcal{R}}\delta f,
\end{equation}
in which $\mathcal{R}=\frac{f_\mathrm{r}}{2M_\mathrm{eff}}$ is the mass responsivity of the resonator.

From Eq. (\ref{eq:mass sensing}), it can be shown that the sensitivity of a resonant sensor is determined by the mass responsivity $\mathcal{R}$ and the minimum detectable frequency shift $\delta f$. It is known that $\mathcal{R}$ is determined by the structure and material properties of the resonator, while the minimum detectable frequency change is determined by the short- and long-term resonance frequency stability of the device that can be evaluated by the $Q$ factor and amplitude of oscillation \cite{ekinci2004ultimate}. It is also worth noting that in the presence of adsobates, the resonance frequency shift $\delta f$ is a convolution of both the stiffness and the mass. However, the stiffness effect comes into play at regions where the resonator undergoes high changes of curvature, close to the nodal lines \cite{ramos2006origin,belardinelli2018modal}. Therefore, by placing an added adsorbate or particle close to an anti-node, this effect can be minimized.

\begin{figure}[h]
	\centering
	\includegraphics[width=12cm]{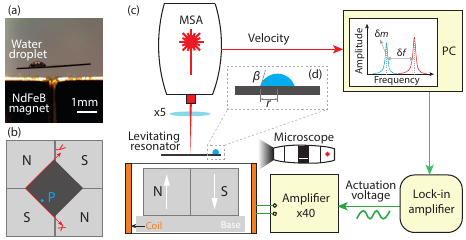}
	\caption{Levitating resonator and experimental setup. (a) Levitating pyrolytic graphite plate above 4 NdFeB magnets with alternating magnetization. A water droplet is dispensed on top of the graphite plate; (b) Top view of the levitating plate (not to scale) shows the position (point P) of the droplet, where N and S stand for north and south pole of the magnet, respectively; (c) Schematic of the sensing principle and the measurement setup comprising a Polytec LDV, an electromagnetic excitation coil, a lock-in amplifier for actuation and readout and, a USB microscope to measure the dimensions of the droplet. The actuation voltage is amplified by a $40\times$ voltage amplifier that drives the levitating plate into resonance via a coil; (d) Schematic of a sessile liquid droplet as captured by the in-plane microscope on the surface of the graphite plate (not to scale), where $\beta$ is the contact angle and $r$ is the contact radius.}
	\label{fig:setup}
\end{figure} 
\section{Experiments}
\label{sec: experiment}

Experiments are performed on levitating pyrolytic graphite plates, which are used to measure the mass of glass beads or liquid droplets that are placed on their surface (Fig. \ref{fig:setup}a). The graphite is purchased from MTI Corporation and is cut using a micro laser cutter to obtain $10\times 10 \times 0.24$ mm$^3$ plates. The surface of the plate is polished using a sand paper with \SI{5}{\um} grain to improve the reflected optical readout signal.
In order to levitate the plates, a checkerboard arrangement of 4 cubic permanent NdFeB magnets with alternating out-of-plane magnetization is used (Fig. \ref{fig:setup}b). The remanent magnetic flux density of the magnets is $B_r=\SI{1.4}{T}$. 

To determine the frequency response of the graphite plate, we use an experimental setup that comprises a Polytec MSA-400 laser Doppler vibrometer (LDV) to measure the out-of-plane velocity of the plate and a Zurich UHFLI lock-in amplifier to drive the levitating plate into resonance through a dedicated electromagnetic coil (Fig. \ref{fig:setup}c) that is surrounding the permanent magnets. The current through the coil generates an alternating magnetic field in the vertical direction that actuates the permanent magnets. The motion of the magnets modulates the magnetic field on the plate and brings it into resonance. Although there is also a direct electromagnetic force on the plate from the coil, this force is estimated to be significantly smaller than the forces generated by the motion of the permanent magnets. To detect the velocity of the resonator, the LDV laser beam is focused on the plate surface and the vibrometer signal obtained by the MSA laser head is  analyzed using the Polytec decoder. Next, the acquired velocity is transferred to the Polytec PSV software for frequency response analysis and obtaining the resonance frequencies and $Q$ factor, or the lock-in amplifier for evaluating the frequency stability in closed loop. To accurately measure the dimensions of the droplet, an in-plane oriented microscope is also used in our setup (Fig. \ref{fig:setup}a,d). The microscope is used to trace the variations in the contact angle $\beta$ and contact radius $r$ of the liquid droplet throughout evaporation. All our experiments are conducted at atmospheric conditions and at room temperature. 


\section{Results and discussion}
\subsection{Dynamic characterization}
\begin{figure}
	\centering
	\includegraphics[height=5cm]{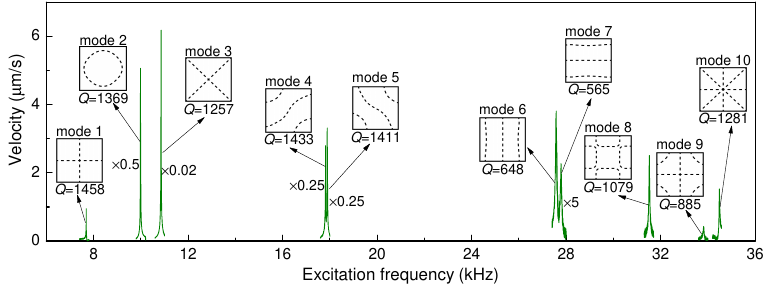}
	\caption{Experimental resonance peaks, Q-factors and mode shapes, as experimentally determined by the LDV, of 10 bending modes of the graphite plate with dimensions of $10\times10\times\SI{0.24}{mm}$, levitating above 4 cubic NdFeB magnets with side length of \SI{12}{mm}. The dashed lines indicate the nodal lines of the measured modes. Peaks have been rescaled by the indicated factors, and $Q$ factors determined by fits. In particular mode 3 has a high amplitude, which is attributed to its efficient actuation by the magnet configuration (Fig. \ref{fig:setup}b).
	}
	\label{fig:dynamics characterization}
\end{figure} 
To characterize the resonant response of the levitating plate, we place the bare plate on the magnets and excite the plate with a periodic chirp signal over a large frequency range. Once we identify a resonance peak,  we perform a narrow-band frequency sweep around the resonance frequency while positioning the laser on the anti-node (point of maximum amplitude) of the excited mode. The outcome of this procedure is shown in Fig. \ref{fig:dynamics characterization} where the first 10 resonance modes of the levitating plate are found. In Fig. \ref{fig:dynamics characterization}, for each resonance peak a schematic of the corresponding mode shape as determined by the LDV is also shown.
To extract the quality factor associated with the modes, we fit a Lorentzian to each resonance peak and obtain $Q \approx 1500$ for the first five modes of vibration in air, which is much higher than the $Q$ of the levitating plate's rigid body modes \cite{d92503966a0041f9a349abdf1003f35a}. It is noted that the $Q$ of the resonator is a result of three main energy dissipation sources: material damping, air damping and eddy current damping. The relatively high $Q$s observed in air make our graphite resonator an interesting candidate for mass sensing applications, where high $Q$s are beneficial for improving detection limits \cite{ekinci2004ultimate}. 

\subsection{Mass responsivity of the resonator}

\label{sec: working principle}
To calibrate the responsivity of the resonant mass sensor, small glass beads with mass $\delta m$ are placed on the graphite plate. This causes a change in the resonance frequency  $\delta f$ associated with a bending mode. We choose to perform our mass measurements using the 3$^{rd}$ bending mode since it has the highest amplitude of all modes of vibration (see Fig. \ref{fig:dynamics characterization}), i.e., this mode has the highest signal-to-noise ratio. We attribute the high amplitude of the third mode mainly to the high efficiency actuation due to the resemblance of the mode shape (Fig. \ref{fig:dynamics characterization}) to the magnet configuration(see Fig. \ref{fig:setup}b), while noting that the direction of the force of the coil on the magnets depends on their respective magnetization direction.

\begin{figure}
	\centering
	\subfigure{\includegraphics[height=4.5cm]{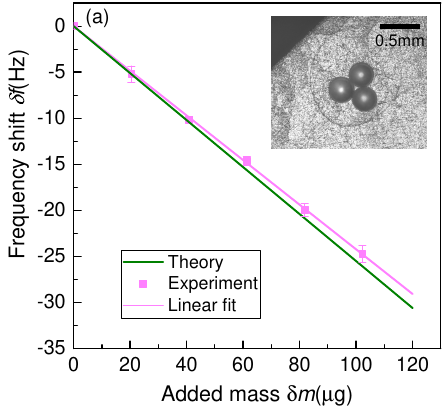}}
	\subfigure{\includegraphics[height=4.5cm]{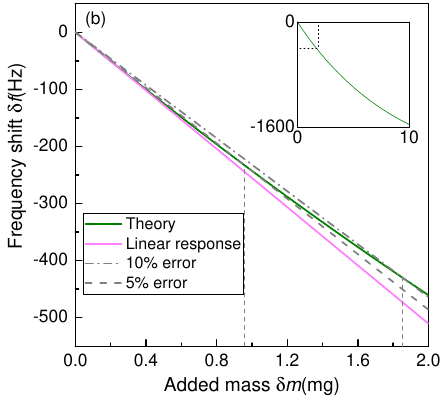}}
	\subfigure{\includegraphics[height=4.7cm]{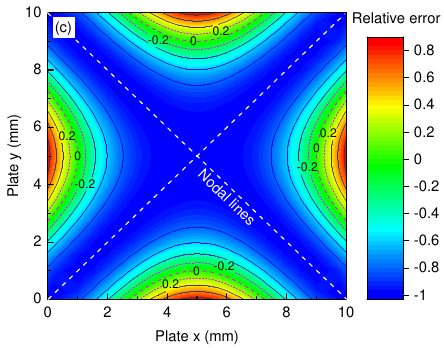}}
	\caption{Mass responsivity and error analysis. (a) Frequency shift as a function of the mass of the added glass beads, as obtained from experiments and as estimated using Rayleigh-Ritz method. The inset shows an image of three glass beads on top of the levitating plate, taken with a microscope; (b) Frequency shift as a function of the added mass obtained numerically, showing a linear relationship between the frequency shift and added mass; (c) Relative error ($\frac{\delta m_\mathrm{P}-\delta m}{\delta m}$), where $\delta m$ is the actual added mass and $\delta m_\mathrm{P}$ is the mass calculated from Eq. (\ref{eq:mass sensing}) using the mass responsivity of point P ($\mathcal{R}_\mathrm{P}$).
	}
	\label{fig: mass responsivity}
\end{figure}

Eq. (\ref{eq:mass sensing}) states that in order to obtain the added mass $\delta m$, it is essential to first determine the responsivity of the sensor. Here, we use glass beads of diameter \SI{250}{\um} with known density to obtain the experimental mass responsivity $\mathcal{R}_\mathrm{exp}$. The glass beads are placed using water droplets created by syringe near the anti-node of the 3$^{rd}$ bending mode (point P: $x=\SI{5}{mm}, y=\SI{1}{mm}$ in Fig. \ref{fig: mass responsivity}c and Fig. \ref{fig:setup}b). To add the glass beads, the graphite plate is removed from the experimental setup, and is then placed back to measure the resonance frequency in the presence of added particles. We note that since the levitation system does not require any clamping, no calibration is required every time the resonator is placed back on top of the magnets. The frequency shift of the resonator after adding different numbers of glass beads is plotted in Fig. \ref{fig: mass responsivity}a. A linear relationship is apparent between the added mass and the frequency shift,  with a slope of \SI{-0.24}{Hz/\ug} that is the experimental mass responsivity  $\mathcal{R}_\mathrm{exp}$ of our resonator, from which $M_\mathrm{eff,exp}=\SI{2.24e-5}{kg}$. 

In order to evaluate the accuracy of our measurements, we construct a numerical model of the levitating plate using the Rayleigh-Ritz method \cite{alijani2013theory,alijani2016damping} and calculate the resonance frequency shifts of the plate with added mass (details of the numerical modelling can be found in Appendix A). Using the numerical model, we determine the frequency shift $\delta f$ as a function of add mass $\delta m$ which is placed on point P.
Our numerical results are also included in Fig. \ref{fig: mass responsivity}a obtaining a theoretical mass responsivity $\mathcal{R}_\mathrm{th}=\SI{-0.25}{Hz/\ug}$ in good agreement with the experimental value. Using the numerical model, we can estimate the range of  $\delta m$ over which the shift in the resonance frequency $\delta f$ varies linearly with the added mass. For this, we perform simulations over a large number of added mass increments and find that when the added mass is below \SI{0.96}{mg}, the relative error between the real  and the estimated mass using Eq. (\ref{eq:mass sensing}) is less than \SI{5}{\percent} (see Fig.\ref{fig: mass responsivity}b). This result suggests that we can use Eq. (1) with confidence to measure mass changes smaller than \SI{1}{mg}. We note that this value is much smaller than the bearing capacity of the levitating plate which is estimated to be \SI{227}{mg} (4 times the plate's mass).

In addition, when using bending vibrations for mass sensing, the position where the mass is placed on the resonator can influence the accuracy of the mass responsivity \cite{ramos2006origin}. To investigate how the position of the added mass affects the accuracy of our measurements, we use the same model to estimate the frequency shift associated with a certain mass (\SI{10}{\ug}) added at different locations on the plate. The result of this numerical study is shown in Fig. \ref{fig: mass responsivity}c, and highlights the error that can be induced in determining $\delta m$, if the added mass is placed on different locations. The relative error can be expressed as $\frac{\delta m_\mathrm{P}-\delta m}{\delta m}$, where $\delta m$ is the actual mass of the particle and $\delta m_\mathrm{P}$ is the mass determined from Eq. (\ref{eq:mass sensing}) using the mass responsivity of point P ($\mathcal{R}_\mathrm{P}$). It could be seen that putting the point mass \SI{0.3}{mm} away from point P ($x=\SI{5}{mm}, y=\SI{0.7}{mm}$) results in an error of \SI{20}{\percent}, indicating that the sensor is pretty sensitive to the placement. Also, this error grows when placing the added mass further away from P and closer to the nodal lines. Therefore, the choice of vibration mode to be used for sensing, depends on the number and distribution of nodal lines associated with that mode. Among the identified mode shapes shown in Fig. \ref{fig:dynamics characterization}, the first three bending modes have the lowest number of nodal lines while exhibiting a decent $Q$. Therefore, they are ideal resonances for investigating the mass sensing concept, especially for droplets which require a certain amount of contact area.

\subsection{Measuring liquid density and evaporation rate}
\begin{figure}
	\centering
	\subfigure{\includegraphics[height=6cm]{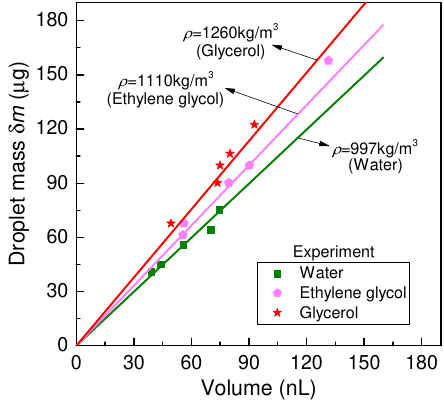}}
	\caption{Plot of the estimated droplet mass from the change in the resonance frequency of the graphite plate, versus droplet volume determined from the microscope image for droplets of water, ethylene glycol and glycerol placed near the antinode position P.
	}
	\label{fig: density}
\end{figure}
Measuring the mass and density of liquid droplets has many applications in bio/chemical technologies \cite{alvarez2010microcantilever,gfeller2005rapid,waggoner2007micro,kumar2011bifurcation}. Here, we measure the density of different liquids and evaporation rate (mass change over time) of \si{nL} volume droplets using our calibrated levitating graphite plate. 
To measure liquid density, we first dispense droplets of different liquids and volume using a syringe near the anti-node of the resonator (point P in Fig 1b) and detect the frequency shift due to the added mass. Next, using the experimental mass responsivity $\mathcal{R}_\mathrm{exp}$ determined in Section 4.2, we estimate  $\delta m$ associated with the dispensed droplets. At the same time, using an in-situ microscope and an image processing code implemented in Matlab (details given in Appendix B), we measure the volume of the droplets by estimating their contact radius $r$ and contact angle $\beta$ (Fig. \ref{fig:setup}d). 
In Fig. \ref{fig: density} we report the mass of droplets of water, ethylene glycol, and glycerol of different volumes, measured using our resonator. It can be observed that the measured values follow the theoretical curves for known densities of these liquids. 

\begin{figure}[h]
	\centering
	\subfigure{\includegraphics[height=4.5cm]{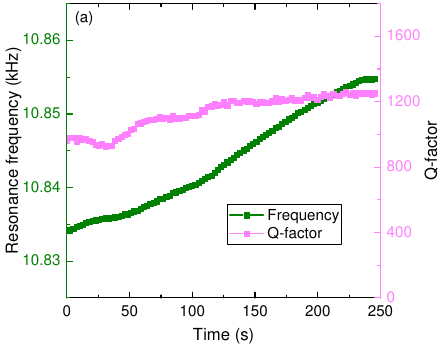}}
	\subfigure{\includegraphics[height=4.5cm]{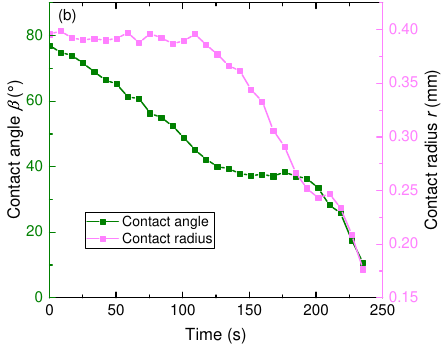}}
	\subfigure{\includegraphics[height=4.5cm]{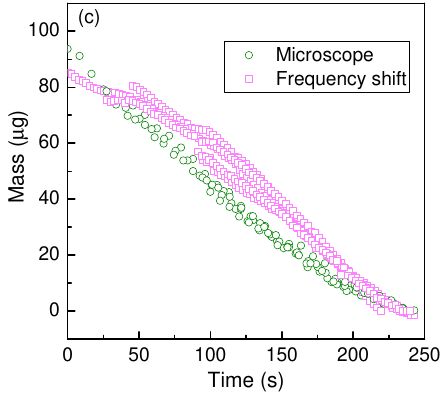}}
	\caption{Measuring evaporation rate of water droplets. (a) Resonance frequency (green) and Q-factor (pink) as a function of time while the water droplet is evaporating, measured using the setup shown in Fig. \ref{fig:setup}c; (b) Contact angle (green) and contact radius (pink) measurements during evaporation; (c) Measured mass of 4 water droplets with different initial volume as a function of time, obtained using both the in-situ microscope (green) and frequency shift method (pink).
	}
	\label{fig: evaporation}
\end{figure}

To investigate the applicability of our levitating resonator for real-time sensing, we next use the resonant levitating mass sensor for detecting the evaporation rate of water droplets using resonance frequency shifts. We dispense droplets of different volumes, and estimate evaporation rate as $\dot {m}=\delta m/\delta t$, where $\delta m$ is the mass change over time $\delta t$ as evaluated using Eq. (\ref{eq:mass sensing}). Fig. \ref{fig: evaporation}a shows the real-time change of resonance frequency and $Q$ of the resonator during the evaporation of a \SI{94}{nL} water droplet. An increase in the resonance frequency of the resonator is observed consistent with a decrease in the volume of the droplet. It is seen that the rate $df_r/dt$ increases after $120s$ and becomes zero beyond $250 s$ confirming complete evaporation of the droplet. 
In Fig. \ref{fig: evaporation}a, the $Q$ of the resonance mode in the presence of the droplet is around $\sim 1000$ and increases to around 1200 after evaporation. Therefore, the influence of the change of $Q$ on the observed resonance frequency shift is found to be negligible during evaporation since $f_\mathrm{r}=f_\mathrm{u}\sqrt{1-(\frac{1}{2Q})^2}$, where $f_\mathrm{u}$ is the undamped resonance frequency. It should also be noted that surface interactions between the droplet might affect the stiffness $K_{\rm eff}$ and might be extracted separately by monitoring both amplitude and frequency of the resonator. 

To compare our results to a second method for estimating mass change from volume change assuming constant density, we also use the in-situ microscope of the setup to monitor the volumetric changes of the water droplet. In Fig. \ref{fig: evaporation}b we show the variation of droplet's contact angle $\beta$ and contact radius $r$ during evaporation using this second method (see Appendix B). It can be seen that the droplet initially evaporates in the constant contact radius mode until $120s$, after which $r$ starts to decrease linearly in time. This decrease is consistent with the increase in $df_r/dt$ that we observed in Fig. \ref{fig: evaporation}a, since the resonance frequency is influenced by the distribution of the added mass (see Fig. \ref{fig: mass responsivity}c). By detecting $\beta$ and $r$ throughout evaporation, we can now calculate the volume changes and, accordingly, the mass of the water droplet using this second technique ($m=\frac{\pi\rho}{3}(\frac{r}{\sin\beta})^3(2+\cos\beta)(1-\cos\beta)^2$ \cite{polyanin2006handbook}). 

We repeat the same experiment for 4 water droplets of nL volumes and compare the mass estimations by both methods. These measurements are summarized in Fig. \ref{fig: evaporation}c where it can be observed that the estimated mass using our resonator is close to the one obtained by tracing volumetric changes using the microscope, with differences attributed to measurement errors. 

\subsection{Frequency stability}
The mass precision of resonant mass sensors depends on the resolution with which frequency changes due to added particles/droplets can be determined. Therefore, the mass resolution or the minimum detectable particle mass is influenced not only by the responsivity  $\mathcal{R}$ of the resonant sensor, but also by its frequency stability. 
In other words,  for a mass change $\delta m$ to be detected by the frequency change $\delta f$, it is essential that the shift in the frequency is greater than the frequency imprecision $\sigma_a$ of the resonator \cite{sansa2016frequency}. This frequency imprecision depends on the time $\tau$ over which a frequency measurement is averaged, and can be defined as the Allan deviation of consecutive resonance frequency measurements each averaged over a time period $\tau$ (gate time) as follows:
\begin{equation}
    \sigma_a=\sqrt{\frac{1}{2 N} \sum_{n=0}^{N-1} \left( \frac{\bar f_{n+1}-\bar f_{n}}{f_0}\right)^2} ,
    \label{eq:allan}
\end{equation}
in which $\bar f_{n}$ is the time average of the frequency measurement during the $n^{th}$ gate time within a total $N$ intervals and $f_0$ is the mean frequency calculated over the frequency tracking operation. To measure the frequency stability and thus Allan deviation of our levitating resonator, we use the PLL of the Zurich UHFLI lock-in amplifier and operate our measurements in closed loop. The PLL uses the output phase of the resonator's $3^{rd}$ bending mode to control the excitation frequency applied to the coil using a PI controller with proportional constant $k_p$ and integral constant $k_i$. To implement the PLL, we use constants $k_p$ =\SI{-1.4}{Hz/deg} and $k_i$ = \SI{-2.8}{Hz/deg/s} that
ensure a stable closed-loop operation and establish a PLL bandwidth of \SI{111.6}{Hz}.
Moreover, to maintain the resonator close to resonance, we set the phase set-point to $\pi/2$ after correcting for the shift introduced by the equipment.   
\begin{figure}
	\centering
	\subfigure{\includegraphics[height=5cm]{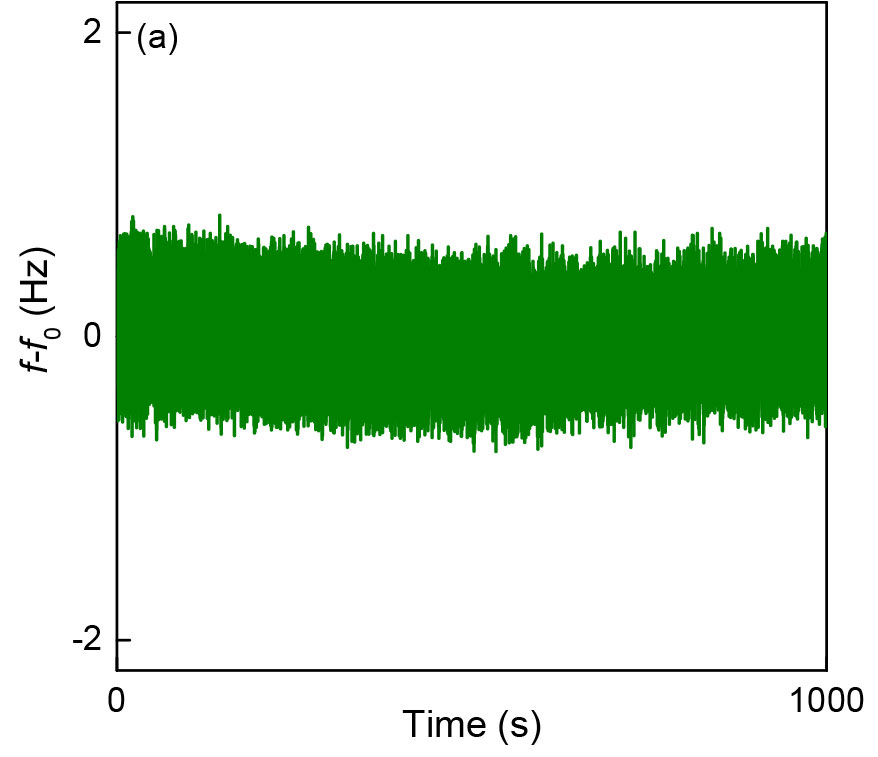}}
	\subfigure{\includegraphics[height=5cm]{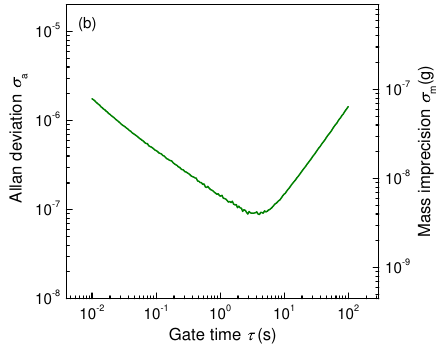}}
	\caption{Frequency stability measurement. (a)Frequency fluctuation as a function of time; (b)Allan deviation and mass imprecision as a function of gate time of the levitating plate measured using a LDV and lock-in amplifier.
	}
	\label{fig: allan deviation}
\end{figure}

 Fig. \ref{fig: allan deviation}a shows the recorded resonance frequency data in a time span of \SI{1000}{s} and Fig. \ref{fig: allan deviation}b shows its associated Allan deviation $\sigma_a$. We observe that the Allan deviation is approximately proportional to  $\tau^{-1/2}$ for short gate times, which can be attributed to white noise sources, like from thermomechanical noise or from the LDV and actuation system  \cite{manzaneque2020method}. For long gate times, the Allan deviation is proportional to $\tau$ which might be attributed to drift. 
 Using the measured Allan deviation and experimental mass responsivity, we evaluate the mass imprecision of our resonant sensor  ($\sigma_\mathrm{m}=2M_\mathrm{eff,exp}\sigma_a$\cite{manzaneque2020method} ), and report the values in Fig. \ref{fig: allan deviation}b. The minimum mass imprecision is found to be  \SI{4.0}{ng} at $\tau=\SI{3}{s}$, which shows the potential of our resonant weighing scale for detecting mass changes due to particles down to a few $\si{ng}$, for example the mass fluctuation in cells \cite{martinez2017inertial}.

\section{Conclusions}
In summary, we demonstrate a diamagnetically levitating graphite mass sensor that is electromagnetically driven into resonance. We use a laser Doppler vibrometer to characterize the first 10 elastic modes of the resonator and use glass beads with known mass to calibrate the mass responsivity of the levitating plate. By dispensing nano-liter droplets of different liquids and tracking the resonance frequency changes, we show the potential of the weighing scale to measure small mass changes due to the evaporation of the liquid. Finally, by operating the resonator in closed loop and measuring the Allan deviation of the frequency fluctuations, we show that mass resolutions down to a few nano-gram are within reach with low cost millimeter scale diamagnetically levitating sensors, showing the potential of these resonators for biological and chemical sensing applications.

\section*{Acknowledgements}
This work was carried out under the 17FUN05 PhotOQuanT project, which has received funding from the EMPIR program, co-financed by the Participating States and the European Union’s Horizon 2020 research and innovation program. This work has also received funding from ERC starting grant ENIGMA (802093) and Graphene Flagship (881603). X.C acknowledges financial support from China Scholarship Council.

\section*{Appendix A: Rayleigh-Ritz method}
\label{sec: elastic modelling}
\setcounter{figure}{0} 
\begin{figure}[h]
	\centering
	\renewcommand{\thefigure}{A\arabic{figure}}
	\includegraphics[width=12cm]{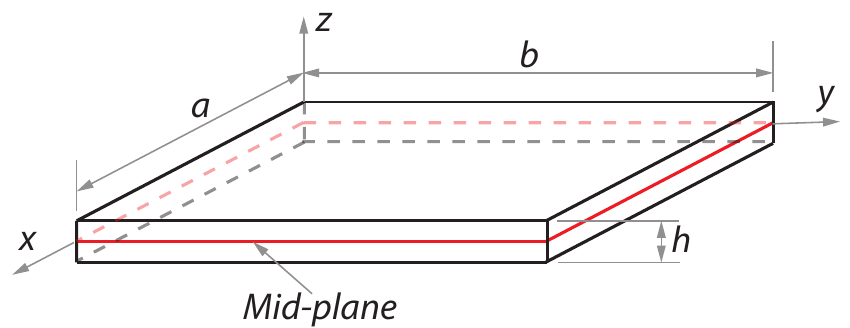}
	\caption{Geometry of the rectangle plate and coordinate system}
	\label{fig: geometry of plate}
\end{figure}
A rectangular plate with in-plane dimensions $a$ and $b$ and thickness $h$ is considered in an orthogonal coordinate system $(O: x, y, z)$, as shown in \cref{fig: geometry of plate}. Three displacements $u, v$ and $w$ are used to describe the plate middle surface deformation.
The strain-displacement relations of the plate can be written as:
\begin{equation}
\setcounter{equation}{1}
\renewcommand{\theequation}{A\arabic{equation}}
\begin{aligned}
\epsilon_{x}&=\epsilon_{x_0}+zk_x,\\
\epsilon_{y}&=\epsilon_{y_0}+zk_y,\\
\gamma_{xy}&=\gamma_{xy_0}+zk_{xy},
\end{aligned}
\end{equation}
where
\begin{equation}
\renewcommand{\theequation}{A\arabic{equation}}
\begin{aligned}
	\epsilon_{x_0}&=\frac{\partial u}{\partial x},\\
	\epsilon_{y_0}&=\frac{\partial v}{\partial y},\\
	\gamma_{xy_0}&=\frac{\partial v}{\partial x}+\frac{\partial u}{\partial y},\\
	k_x&=-\frac{\partial ^2w}{\partial x^2},\\
	k_y&=-\frac{\partial ^2w}{\partial y^2},\\
	k_{xy}&=-\frac{\partial ^2w}{\partial x\partial y}.
\end{aligned}
\end{equation}
The kinetic energy of the plate is obtained as:
\begin{equation}
\renewcommand{\theequation}{A\arabic{equation}}
T_\mathrm{p}=\frac{1}{2}\rho h\int_{0}^{a}\int_{0}^{b}(\dot{u}^2+\dot{v}^2+\dot{w}^2)\mathrm{d}x\mathrm{d}y+\frac{1}{2}m_\mathrm{add}(\dot{u} (x_a,y_a)^2+\dot{v} (x_a,y_a)^2+\dot{w} (x_a,y_a)^2),
\end{equation}
where the overdot indicates differentiation with respect to time, and $m_\mathrm{add}$ is the mass of the added particle. Moreover,  $x_a$ and $y_a$ are the $x,y$ coordinates of the point where the particle is attached.

The elastic strain energy of the plate assuming isotropic material properties is \cite{alijani2013theory}:
\begin{equation}
\renewcommand{\theequation}{A\arabic{equation}}
\label{eq: T and U}
\begin{aligned}
U_\mathrm{p}&=\frac{1}{2}\frac{Eh}{1-\nu^2}\int_{0}^{a}\int_{0}^{b}(\epsilon _{x_0}^2+\epsilon _{y_0}^2+2\nu\epsilon_{x_0}\epsilon_{y_0}+\frac{1-\nu}{2}\gamma_{xy_0}^2)\mathrm{d}x\mathrm{d}y\\
&+\frac{1}{2}\frac{Eh^3}{12(1-\nu^2)}\int_{0}^{a}\int_{0}^{b}(k _{x}^2+k _{y}^2+2\nu k_{x}k_{y}+\frac{1-\nu}{2}k_{xy}^2)\mathrm{d}x\mathrm{d}y,\\
\end{aligned}	
\end{equation}
where $E$ is the Young's modulus and $\nu$ is the Poisson's ratio. In order to obtain natural frequencies and modes of vibration of the plate, we assume synchronous motion and express displacements $u,v$ and $w$ in the follwoing form:
\begin{equation}
    \renewcommand{\theequation}{A\arabic{equation}}
	\begin{aligned}
	u(x,y,t)&=U(x,y)g(t),\\
	v(x,y,t)&=V(x,y)g(t),\\
	w(x,y,t)&=W(x,y)g(t),
	\end{aligned}
\end{equation}
where the functions $U(x,y), V(x,y)$ and $W(x,y)$ represent the mode shapes, while $g(t)$ is harmonic time function and is the same for all displacements. To  obtain mode of vibration we use power polynomials:
\begin{equation}
\renewcommand{\theequation}{A\arabic{equation}}
\begin{aligned}
\label{eq:trial equation}
U(x,y)=\sum_{m=0}^{M}\sum_{n=0}^{N}a_{mn} x^m y^n,\\
V(x,y)=\sum_{m=0}^{M}\sum_{n=0}^{N}b_{mn} x^m y^n,\\
W(x,y)=\sum_{m=0}^{M}\sum_{n=0}^{N}c_{mn} x^m y^n.\\
\end{aligned}		
\end{equation}
In order to find the linear free vibration response, a vector $\boldsymbol{\mathrm{p}}$ comprising all unknown coefficients of Eq. A\ref{eq:trial equation} is built as follows:
\begin{equation}
    \renewcommand{\theequation}{A\arabic{equation}}
	\boldsymbol{\mathrm{p}}=\left\lbrace a_{00},...,a_{mn},b_{00},...,b_{mn},c_{00},...,c_{mn}\right\rbrace.  
\end{equation}
The dimension of vector $\boldsymbol{\mathrm{p}}$ is $N_{max}$ that is equal to the total number of unknowns in Eq. A\ref{eq:trial equation}. Therefore, $N_{max}=3(M+1)(N+1)$. Next, the Lagrange equations of motion are obtained by assuming $\mathrm{g}(t)=\boldsymbol{\mathrm{p}} g(t)$ where $g(t)=e^{j\omega t}$ with $j$ being the imaginary unit and $\omega$ the vibration frequency as follows:
\begin{equation}
\renewcommand{\theequation}{A\arabic{equation}}
\label{eq:Lagrange}
\frac{\mathrm{d}}{\mathrm{d}t}\frac{\partial T_p}{\partial\dot{g}_k}-\frac{\partial T_p}{\partial g_k}+\frac{\partial U_p}{\partial g_k}=0, k=1,..., N_{max},		
\end{equation}
which results in 
\begin{equation}
    \renewcommand{\theequation}{A\arabic{equation}}
	\label{eq:eigenvalue}
	(-\omega ^2\boldsymbol{\mathrm{M}}+\boldsymbol{\mathrm{K}})\boldsymbol{\mathrm{p}}=0,
\end{equation}
where $\boldsymbol{\mathrm{M}}$ is the mass matrix and $\boldsymbol{\mathrm{K}}$ is the stiffness matrix of dimension $N_{max}\times N_{max}$.
Setting the determinant of the vector $\boldsymbol{\mathrm{p}}$ equal to zero, the eigenvalues will be obtained and substituting each eigenvalue into Eq. A\ref{eq:eigenvalue} gives its corresponding eigenvector. In order to obtain the natural mode corresponding to the $i_{th}$ eigenvector, we substitute the coefficients $a_{mn}, b_{mn}$, and $c_{mn}$ with $a_{mn}^i, b_{mn}^i$ and $c_{mn}^i$ which are  the components of the $i-\mathrm{th}$ eigenvector obtained from Eq. A\ref{eq:trial equation} as:
\begin{equation}
\renewcommand{\theequation}{A\arabic{equation}}
\begin{aligned}
\label{eq:i-th mode shape}
U^{(i)}(x,y)&=\sum_{m=0}^{M}\sum_{n=0}^{N}a^{(i)}_{mn} x^m y^n,\\
V^{(i)}(x,y)&=\sum_{m=0}^{M}\sum_{n=0}^{N}b{(i)}_{mn} x^m y^n,\\
W^{(i)}(x,y)&=\sum_{m=0}^{M}\sum_{n=0}^{N}c{(i)}_{mn} x^m y^n.\\
\end{aligned}		
\end{equation}
It is worth to note that in order to avoid matrix ill-conditioning in using power polynomials, high numerical accuracy is required. In this work  the software $\emph{Mathematica 11.3}$ has been used with 300
digits of accuracy to avoid matrix ill-conditioning. 

Using this model, by comparing the estimated resonance frequencies and the experimental resonance frequencies from Fig. 2 of the paper, we find that the graphite plate has an experimentally estimated Young's modulus $E=\SI{3.72e10}{Pa}$ and Poisson's ratio $\nu=-0.25$ in reasonable agreement with literature values \cite{garber1963pyrolytic}. After that, by changing the $m_\mathrm{add}$, we can calculate the frequency shift $\delta f$, thus calculating the mass responsivity of the third mode at point P to be $\mathcal{R}_\mathrm{th}=\SI{-0.25}{Hz/\ug}$.

\section*{Appendix B: Image processing method}
Typically, a liquid droplet can be viewed as a spherical cap when the contact radius of the droplet is smaller than the capillary length \cite{harth2012simple,cazabat2010evaporation,hu2002evaporation}. The capillary length $l_\mathrm{c}$ of a droplet can be obtained as
\begin{equation}
    \centering
    \label{eq:capillary length}
    l_\mathrm{c}=\sqrt{\frac{\sigma}{\rho g}},
\end{equation}
where $\sigma$ and $\rho $ are the surface tension and density of the liquid, respectively, and $g$ is the gravitational constant. For a water droplet, the capillary length is  $\SI{2.7}{mm}$ which is much larger than the maximum contact radius of the droplet measured in our experiments i.e. $r=\SI{0.4}{mm}$.  The same holds true for our measurements with ethylene glycol  ($r=\SI{0.5}{mm}$) and glycerol ($r=\SI{0.4}{mm}$) whose capillary lengths are $\SI{2.1}{mm}$ and $\SI{2.3}{mm}$, respectively. Therefore, the droplets in our experiments can be seen as spherical caps, and their side view can be fitted with a circle.

After taking an image of the sessile droplet on the surface of graphite plate, we extract the contact radius $r$ and contact angle $\beta$ using an image processing method using Matlab. We first convert the image taken by the microscope (Fig. B1a) into grey scale (Fig. B1b) to make the edges of the droplet and plate more clear. Next, the grey picture is converted to black and white (Fig. B1c), and  from this  image, droplet and plate edges are identified ( Fig. B1d). Finally to obtain the contact radius and contact angle,  a line is drawn at the edge of the plate and a circle is fitted using three points on the edge of the droplet (see Fig. B1b). With the fitted circle and the straight line, the contact angle and radius are calculated in pixels. Using the known thickness of the plate as a reference, these values are then converted to mm.
\setcounter{figure}{0} 
\begin{figure}[h]
	\centering
	\renewcommand{\thefigure}{B\arabic{figure}}
	\includegraphics[width=12cm]{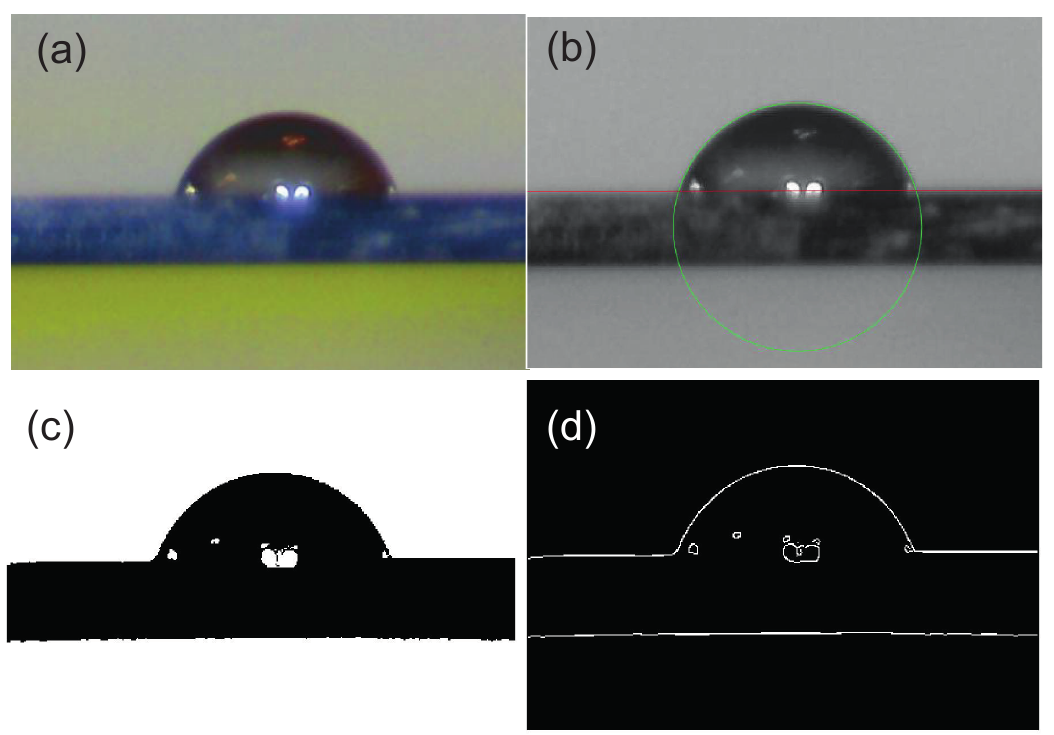}
	\caption{Image processing method. (a)Image of a droplet on top of the plate; (b) Gray scale image and a fitted circle and straight line; (c) Black and white image; (d) Edges of the droplet and plate.}
	\label{fig: image processing}
\end{figure}

\pagebreak

\bibliographystyle{ieeetr}
\bibliography{ref_mass_sensing}

\begin{thebibliography}{10}

\bibitem{jensen2008atomic}
K.~Jensen, K.~Kim, and A.~Zettl, ``An atomic-resolution nanomechanical mass
  sensor,'' {\em Nature nanotechnology}, vol.~3, no.~9, p.~533, 2008.

\bibitem{chaste2012nanomechanical}
J.~Chaste, A.~Eichler, J.~Moser, G.~Ceballos, R.~Rurali, and A.~Bachtold, ``A
  nanomechanical mass sensor with yoctogram resolution,'' {\em Nature
  nanotechnology}, vol.~7, no.~5, pp.~301--304, 2012.

\bibitem{yang2006zeptogram}
Y.-T. Yang, C.~Callegari, X.~Feng, K.~L. Ekinci, and M.~L. Roukes,
  ``Zeptogram-scale nanomechanical mass sensing,'' {\em Nano letters}, vol.~6,
  no.~4, pp.~583--586, 2006.

\bibitem{zhang2014detecting}
Y.~Zhang, ``Detecting the stiffness and mass of biochemical adsorbates by a
  resonator sensor,'' {\em Sensors and Actuators B: Chemical}, vol.~202,
  pp.~286--293, 2014.

\bibitem{belardinelli2020second}
P.~Belardinelli, S.~N. Souza, E.~Verlinden, J.~Wei, U.~Staufer, F.~Alijani, and
  M.~K. Ghatkesar, ``Second flexural and torsional modes of vibration in
  suspended microfluidic resonator for liquid density measurements,'' {\em
  Journal of Micromechanics and Microengineering}, vol.~30, no.~5, p.~055003,
  2020.

\bibitem{wingqvist2007shear}
G.~Wingqvist, J.~Bjurstr{\"o}m, L.~Liljeholm, V.~Yantchev, and I.~Katardjiev,
  ``Shear mode aln thin film electro-acoustic resonant sensor operation in
  viscous media,'' {\em Sensors and Actuators B: Chemical}, vol.~123, no.~1,
  pp.~466--473, 2007.

\bibitem{morten1992resonant}
B.~Morten, G.~De~Cicco, and M.~Prudenziati, ``Resonant pressure sensor based on
  piezoelectric properties of ferroelectric thick films,'' {\em Sensors and
  Actuators A: Physical}, vol.~31, no.~1-3, pp.~153--158, 1992.

\bibitem{waggoner2007micro}
P.~S. Waggoner and H.~G. Craighead, ``Micro-and nanomechanical sensors for
  environmental, chemical, and biological detection,'' {\em Lab on a Chip},
  vol.~7, no.~10, pp.~1238--1255, 2007.

\bibitem{johnson2012biosensing}
B.~N. Johnson and R.~Mutharasan, ``Biosensing using dynamic-mode cantilever
  sensors: A review,'' {\em Biosensors and bioelectronics}, vol.~32, no.~1,
  pp.~1--18, 2012.

\bibitem{cooper2009label}
M.~A. Cooper, {\em Label-free biosensors: techniques and applications}.
\newblock Cambridge University Press, 2009.

\bibitem{schmid2016fundamentals}
S.~Schmid, L.~G. Villanueva, and M.~L. Roukes, {\em Fundamentals of
  nanomechanical resonators}, vol.~49.
\newblock Springer, 2016.

\bibitem{norte2016mechanical}
R.~A. Norte, J.~P. Moura, and S.~Gr{\"o}blacher, ``Mechanical resonators for
  quantum optomechanics experiments at room temperature,'' {\em Physical review
  letters}, vol.~116, no.~14, p.~147202, 2016.

\bibitem{wang2011acoustic}
W.~Wang and D.~Weinstein, ``Acoustic bragg reflectors for q-enhancement of
  unreleased mems resonators,'' in {\em 2011 Joint Conference of the IEEE
  International Frequency Control and the European Frequency and Time Forum
  (FCS) Proceedings}, pp.~1--6, IEEE, 2011.

\bibitem{ghadimi2018elastic}
A.~H. Ghadimi, S.~A. Fedorov, N.~J. Engelsen, M.~J. Bereyhi, R.~Schilling,
  D.~J. Wilson, and T.~J. Kippenberg, ``Elastic strain engineering for ultralow
  mechanical dissipation,'' {\em Science}, vol.~360, no.~6390, pp.~764--768,
  2018.

\bibitem{brandt1989levitation}
E.~Brandt, ``Levitation in physics,'' {\em Science}, vol.~243, no.~4889,
  pp.~349--355, 1989.

\bibitem{d92503966a0041f9a349abdf1003f35a}
X.~Chen, A.~Ke{\c s}kekler, F.~Alijani, and P.~Steeneken, ``Rigid body dynamics
  of diamagnetically levitating graphite resonators,'' {\em Applied Physics
  Letters}, vol.~116, no.~24, 2020.

\bibitem{gieseler2012subkelvin}
J.~Gieseler, B.~Deutsch, R.~Quidant, and L.~Novotny, ``Subkelvin parametric
  feedback cooling of a laser-trapped nanoparticle,'' {\em Physical review
  letters}, vol.~109, no.~10, p.~103603, 2012.

\bibitem{simon2000diamagnetic}
M.~Simon and A.~Geim, ``Diamagnetic levitation: Flying frogs and floating
  magnets,'' {\em Journal of applied physics}, vol.~87, no.~9, pp.~6200--6204,
  2000.

\bibitem{chetouani2006diamagnetic}
H.~Chetouani, C.~Jeandey, V.~Haguet, H.~Rostaing, C.~Dieppedale, and G.~Reyne,
  ``Diamagnetic levitation with permanent magnets for contactless guiding and
  trapping of microdroplets and particles in air and liquids,'' {\em IEEE
  Transactions on Magnetics}, vol.~42, no.~10, pp.~3557--3559, 2006.

\bibitem{pigot2008diamagnetic}
C.~Pigot, H.~Chetouani, G.~Poulin, and G.~Reyne, ``Diamagnetic levitation of
  solids at microscale,'' {\em IEEE Transactions on Magnetics}, vol.~44,
  no.~11, pp.~4521--4524, 2008.

\bibitem{garmire2007diamagnetically}
D.~Garmire, H.~Choo, R.~Kant, S.~Govindjee, C.~Sequin, R.~Muller, and
  J.~Demmel, ``Diamagnetically levitated mems accelerometers,'' in {\em
  TRANSDUCERS 2007-2007 International Solid-State Sensors, Actuators and
  Microsystems Conference}, pp.~1203--1206, IEEE, 2007.

\bibitem{palagummi2018bi}
S.~Palagummi and F.~Yuan, ``A bi-stable horizontal diamagnetic levitation based
  low frequency vibration energy harvester,'' {\em Sensors and Actuators A:
  Physical}, vol.~279, pp.~743--752, 2018.

\bibitem{gao2019centrifugal}
Q.-H. Gao, W.-B. Li, H.-X. Zou, H.~Yan, Z.-K. Peng, G.~Meng, and W.-M. Zhang,
  ``A centrifugal magnetic levitation approach for high-reliability density
  measurement,'' {\em Sensors and Actuators B: Chemical}, vol.~287, pp.~64--70,
  2019.

\bibitem{li2006lateral}
Q.~Li, K.-S. Kim, and A.~Rydberg, ``Lateral force calibration of an atomic
  force microscope with a diamagnetic levitation spring system,'' {\em Review
  of scientific instruments}, vol.~77, no.~6, p.~065105, 2006.

\bibitem{ekinci2004ultimate}
K.~Ekinci, Y.~Yang, and M.~Roukes, ``Ultimate limits to inertial mass sensing
  based upon nanoelectromechanical systems,'' {\em Journal of applied physics},
  vol.~95, no.~5, pp.~2682--2689, 2004.

\bibitem{ramos2006origin}
D.~Ramos, J.~Tamayo, J.~Mertens, M.~Calleja, and A.~Zaballos, ``Origin of the
  response of nanomechanical resonators to bacteria adsorption,'' {\em Journal
  of Applied Physics}, vol.~100, p.~106105, 2006.

\bibitem{belardinelli2018modal}
P.~Belardinelli, L.~Hauzer, M.~{\v{S}}i{\v{s}}kins, M.~Ghatkesar, and
  F.~Alijani, ``Modal analysis for density and anisotropic elasticity
  identification of adsorbates on microcantilevers,'' {\em Applied Physics
  Letters}, vol.~113, no.~14, p.~143102, 2018.

\bibitem{alijani2013theory}
F.~Alijani and M.~Amabili, ``Theory and experiments for nonlinear vibrations of
  imperfect rectangular plates with free edges,'' {\em Journal of Sound and
  Vibration}, vol.~332, no.~14, pp.~3564--3588, 2013.

\bibitem{alijani2016damping}
F.~Alijani, M.~Amabili, P.~Balasubramanian, S.~Carra, G.~Ferrari, and
  R.~Garziera, ``Damping for large-amplitude vibrations of plates and curved
  panels, part 1: Modeling and experiments,'' {\em International Journal of
  Non-Linear Mechanics}, vol.~85, pp.~23--40, 2016.

\bibitem{alvarez2010microcantilever}
M.~Alvarez and L.~M. Lechuga, ``Microcantilever-based platforms as biosensing
  tools,'' {\em Analyst}, vol.~135, no.~5, pp.~827--836, 2010.

\bibitem{gfeller2005rapid}
K.~Y. Gfeller, N.~Nugaeva, and M.~Hegner, ``Rapid biosensor for detection of
  antibiotic-selective growth of escherichia coli,'' {\em Applied and
  environmental microbiology}, vol.~71, no.~5, pp.~2626--2631, 2005.

\bibitem{kumar2011bifurcation}
V.~Kumar, J.~W. Boley, Y.~Yang, H.~Ekowaluyo, J.~K. Miller, G.~T.-C. Chiu, and
  J.~F. Rhoads, ``Bifurcation-based mass sensing using
  piezoelectrically-actuated microcantilevers,'' {\em Applied Physics Letters},
  vol.~98, no.~15, p.~153510, 2011.

\bibitem{polyanin2006handbook}
A.~D. Polyanin and A.~V. Manzhirov, {\em Handbook of mathematics for engineers
  and scientists}.
\newblock CRC Press, 2006.

\bibitem{sansa2016frequency}
M.~Sansa, E.~Sage, E.~C. Bullard, M.~G{\'e}ly, T.~Alava, E.~Colinet, A.~K.
  Naik, L.~G. Villanueva, L.~Duraffourg, M.~L. Roukes, {\em et~al.},
  ``Frequency fluctuations in silicon nanoresonators,'' {\em Nature
  nanotechnology}, vol.~11, no.~6, p.~552, 2016.

\bibitem{manzaneque2020method}
T.~Manzaneque, P.~G. Steeneken, F.~Alijani, and M.~K. Ghatkesar, ``Method to
  determine the closed-loop precision of resonant sensors from open-loop
  measurements,'' {\em IEEE Sensors Journal}, vol.~20, no.~23,
  pp.~14262--14272, 2020.

\bibitem{martinez2017inertial}
D.~Mart{\'\i}nez-Mart{\'\i}n, G.~Fl{\"a}schner, B.~Gaub, S.~Martin, R.~Newton,
  C.~Beerli, J.~Mercer, C.~Gerber, and D.~J. M{\"u}ller, ``Inertial picobalance
  reveals fast mass fluctuations in mammalian cells,'' {\em Nature}, vol.~550,
  no.~7677, pp.~500--505, 2017.

\bibitem{garber1963pyrolytic}
A.~Garber, ``Pyrolytic materials for thermal protection systems,'' {\em
  Aerospace Engineering}, vol.~22, no.~1, pp.~126--137, 1963.

\bibitem{harth2012simple}
M.~H{\"a}rth and D.~W. Schubert, ``Simple approach for spreading dynamics of
  polymeric fluids,'' {\em Macromolecular Chemistry and Physics}, vol.~213,
  no.~6, pp.~654--665, 2012.

\bibitem{cazabat2010evaporation}
A.-M. Cazabat and G.~Guena, ``Evaporation of macroscopic sessile droplets,''
  {\em Soft Matter}, vol.~6, no.~12, pp.~2591--2612, 2010.

\bibitem{hu2002evaporation}
H.~Hu and R.~G. Larson, ``Evaporation of a sessile droplet on a substrate,''
  {\em The Journal of Physical Chemistry B}, vol.~106, no.~6, pp.~1334--1344,
  2002.

\end{thebibliography}
\biboptions{compress}


\end{document}